\documentstyle[amssymb,twocolumn,prl,aps]{revtex}

\twocolumn

\begin{document}

\narrowtext

{\large {\bf Quantum Hall effect at low magnetic fields due to
electron-electron interaction}}

\bigskip

In a recent Letter\cite{Hu} Huckestein showed that at real experimental
conditions it is not possible to observe the quantum Hall effect (QHE) at $%
\omega _{c}\tau <1$ predicted by Khmelnitskii \cite{Kh} ($\omega _{c}=eB/m$
is the cyclotron frequency and $\tau $ is the scattering time). The
observation of such a QHE with reentrant Hall plateaus at low magnetic
fields would give support for the global phase diagram \cite{KLZ} of the
integer QHE as obtained from scaling theory \cite{Kh}. We would like to
point out here that in fact the situation is not so hopeless due to the
influence of the electron-electron interaction.

For a bare high-temperature conductivity $\sigma _{xy}^{0} = \sigma_0
\omega_c \tau / [1 + (\omega_c \tau)^2] \leqslant \sigma _{0}/2<1.5e^{2}/h$
with zero-field conductivity $\sigma _{0}=ne^{2}\tau /m<3e^{2}/h$, only one
QHE plateau with $\sigma _{xy}=e^{2}/h$ is possible. For $\sigma
_{0}>3e^{2}/h$ additional QHE plateaus appear and, therefore, a reentrant
QHE structure can occur for $\omega _{c}\tau <1$ \cite{Hu}. For the quantum
values $\sigma_{xy}^{0} = ie^{2}/h$ the localization length 
\begin{equation}
\xi \sim l\exp \left( \pi ^{2}\sigma _{xx}^{0}~^{2}h^{2}/e^{4}\right) ,
\label{xi}
\end{equation}
is huge ($l$ is the elastic mean free path). For $\sigma_{xx}^{0} = \sigma_0
/ [1 + (\omega_c \tau)^2] > \sigma_0/2 > 1.5 e^2/h$ with $\omega_c \tau < 1$%
, $\xi \gtrsim 10^{9}l$ exceeds all rational sizes of samples \cite{Hu}.
Therefore, a sample for which the low-field QHE could be observed due to the
localization of electrons in a 2D system \cite{Kh}, in fact will be metallic
at $\omega _{c}\tau <1$ down to a small field $B_{IM}.$ The value of $B_{IM}$
can be estimated from $\pi \xi _{0}^{2}B_{IM}\sim \Phi _{0}$, where $\Phi
_{0}=h/e$ is the flux quantum and $\xi _{0}\sim l\exp (\pi \sigma
_{0}h/2e^{2})$ is the zero-field localization length which is much smaller
than the localization length at a magnetic field larger than $B_{IM}$.\ For $%
\sigma _{0}=3e^{2}/h$ and $l=3\times 10^{-8}$ m the value $B_{IM}\sim 10^{-4}
$ T is small.

However, in addition to the localization effects for noninteracting
electrons, the electron-electron interaction results in a logarithmic
decrease of the density of states at the Fermi level and of the dissipative
conductivity \cite{AA} 
\begin{equation}
\sigma _{xx}=\sigma _{xx,0}-\lambda e^{2}/\pi h\ln (T_{0}/T)  \label{cc}
\end{equation}
where $\lambda <1$ is the electron-electron interaction constant of order
unity. At low enough temperatures a Coulomb gap should occur with localized
states near the Fermi level. The localization length should be minimal at
the Fermi energy $E_{F}$, increases for $\left| E-E_{F}\right| $ increasing 
\cite{JW}, and becomes much larger than the sample size outside the gap in
accordance with Eq.(\ref{xi}). Since the dissipative conductance vanishes at
zero temperature, then according to Laughlin's gauge argument the Hall
conductance should be quantized \cite{Laugh}. Therefore, $\sigma _{xy}$
should tend to a quantum value with temperature decreasing. Since $\sigma
_{xy}$ attains different quantum values depending on the $\sigma _{xy}^{0}$
value, transition regions should exist for which $\sigma _{xx}$ has a finite
value and $\sigma _{xy}$ is not quantized at zero temperature. In this
scenario there are no levitating extended states between localized states
below the Fermi level, which are necessary for the QHE in the case of
noninteracting electrons \cite{Kh,Laugh2}. All states below the Coulomb gap
down to the mobility edge near the bottom of the band are delocalized.
However, the flow diagram \cite{Kh} and the global phase diagram \cite{KLZ}
of the QHE should be qualitatively valid in this case.

>From Eq.(\ref{cc}) follows that the temperature at which one could expect
the QHE should be of the order of 
\begin{equation}
T_{1}\sim T_{0}\exp (-\pi \sigma _{xx}^{0}h/e^{2}\lambda )
\end{equation}
For $\sigma _{xx}^{0}=1.5e^{2}/h$ and $\lambda =1$, 0.7 and 0.5, the
temperature $T_{1}\approx 1$, $0.1$ and $0.01$~K, correspondingly, for a
reasonable value $T_{0}\approx \hbar /k_{B}\tau \approx 100$ K.

A support of the above presented scenario is given by the recent observation
of the QHE in heavily doped n-GaAs layers at $\omega _{c}\tau >1$\ but with
thicknesses of the samples $d>l$, when the bare electron spectrum is
continuous like for bulk material \cite{MJ}. The QHE is observed for $\sigma
_{xx}^{0}$ up to $2.6e^{2}/h$ with $\xi \sim 10^{29}l$ as calculated from
Eq.(\ref{xi}). For the explanation of the QHE in this case, the
electron-electron interaction has to be taken into account.

Along these lines one could expect a QHE at low magnetic fields in a
two-dimensional disordered system with a sufficient large constant of
interaction $\lambda $ and scattering rate $1/\tau $, at experimentally
obtainable temperatures. An enhanced spin-splitting will be in favor for
such a situation \cite{spin}.

\medskip

S. S. Murzin$^{1,2}$ and A. G. M. Jansen$^{2}$

$^{1}$Institute of Solid State Physics RAS, 142432, Chernogolovka, Moscow
distr., Russia

$^{2}$Grenoble High Magnetic Field Laboratory, MPI-FKF and CNRS, B.P.166,
F-38042, Grenoble Cedex 9, France

\medskip PACS nubers: 73.40 Hm, 71.30+h, 72.15.Rn, 71.55.Jv

\end{document}